\documentclass{JHEP3} 

\usepackage{epsfig,multicol,bbm}
\usepackage{cite}
\usepackage{slashed}

\newcommand\fverb{\setbox\fverbbox=\hbox\bgroup\verb}
\newcommand\fverbdo{\egroup\medskip\noindent%
            \fbox{\unhbox\fverbbox}\ }
\newcommand\fverbit{\egroup\item[\fbox{\unhbox\fverbbox}]}
\newbox\fverbbox

\title{  Probe R-parity violating stop resonance at the LHeC }
\author{Wei Hong-Tang, Zhang Ren-You, Guo Lei, Han Liang,  Ma Wen-Gan, Li Xiao-Peng and Wang Ting-Ting\\
Department of Modern Physics, University of Science and Technology of China, Hefei, Anhui 230026, P.R.China \\
E-mail: \email{weiht@mail.ustc.edu.cn}, \email{zhangry@ustc.edu.cn},
\email{guolei@mail.ustc.edu.cn}, \email{hanl@ustc.edu.cn},
\email{mawg@ustc.edu.cn}, \email{qqpengal@mail.ustc.edu.cn},
\email{tingting@mail.ustc.edu.cn}}

\abstract{ We investigate the possibility of detecting single sqaurk production
at the proposed LHeC collider, in the framework of R-parity violating supersymmetry.
Taking advantage of the enhancement of the direct resonance production of squark and
the distinctive kinematics distributions of $\tilde{q}\rightarrow l q$ two body decay
final states, the LHeC provides excellent opportunities of probing R-violating $\hat{L}\hat{Q}\hat{D}$
interactions at unprecedented level compared to all the knowledge derived from indirect
low energy nucleon measurements. If no apparent deviation from SM
predictions on high invariant mass of muon and b-quark final states at the LHeC with 1$fb^{-1}$ data,
the sensitivities on $\hat{L}\hat{Q}\hat{D}$ coupling constant $\lambda^{'}_{131}
\times \lambda^{'}_{233}$ can be improved by nearly four orders, at energy scale about $100~GeV$. }


\vfill \eject

\begin{document}

\par
\section{Introduction}
In the current structure of the Standard Model (SM), the
conservations of baryon number B and lepton number L are automatic
consequence of the gauge invariance and renormalizability. On the
other hand, neutrino oscillation observations~\cite{neutrino-1,
neutrino-2, neutrino-3} manifest strong lepton flavor violation, and
imply that the SM is not a fundamental particle theory and should be
extended. In minimal supersymmetric (SUSY) extensions of the SM, a
new multiplicative quantum number, $R=(-1)^{2S+L+3B}$, is introduced
in terms of $B$, $L$ and the spin quantum number~$S$, to distinguish
the SM particles ($R=+1$) from their SUSY partners ($R=-1$). In the
most general supersymmetric potential, $R$-parity violating (RPV)
interactions can be included as
\begin{equation}
\label{eqn:RPVpotential} {\cal{W}}_{\slashed{R}_{p}}
=\frac{1}{2}\epsilon_{ab} \lambda_{ijk} \hat{L}^{a}_{i}
\hat{L}^{b}_{j} \hat{E}_{k} + \epsilon_{ab}\lambda^{'}_{ijk}
\hat{L}^{a}_{i} \hat{Q}^{b}_{j} \hat{D}_{k} +
\frac{1}{2}\epsilon_{\alpha\beta\gamma}\lambda^{''}_{ijk}
\hat{U}^{\alpha}_{i} \hat{D}^{\beta}_{j} \hat{D}^{\gamma}_{k} +
\epsilon_{ab} \delta_{i}\mu_{i} \hat{L}^{a}_{i} \hat{H}^{b}_{2}
\end{equation}
where i,j,k=(1,2,3) are generation indices, a,b=(1,2) are SU(2)
isospin indices, and $\alpha,\beta,\gamma$ are SU(3) color indices.
$\hat{L}$ and $\hat{Q}$ are the lepton and quark $SU(2)$ doublet
superfields, $\hat{E}$, $\hat{U}$ and $\hat{D}$ denote the singlets,
and $\hat{H}$ is the Higgs doublet. The bilinear terms
$\hat{L}\hat{H}$ mix the lepton and Higgs superfield with the
Higgsino mass parameter $\mu$, and consequently generate masses of
neutrinos. All the trilinear terms, $\hat{L}\hat{L}\hat{E}$,
$\hat{L}\hat{Q}\hat{D}$ and $\hat{U}\hat{D}\hat{D}$ with
dimensionless R-violating Yukawa couplings $\lambda$, $\lambda^{'}$
and $\lambda^{''}$ respectively, only violate either L- or
B-symmetry alone. Constraints on RPV couplings obtained so far are
well summarized in Ref.~\cite{rpv_report}.

\par
Along with introducing compatible description of neutrino
oscillation in a natural way, the most attracting phenomena of RPV
is to allow single production of SUSY particles. Contrary to the
pair productions of SUSY particles in the R-conserved frame, the RPV
resonance production at colliders would dramatically reduce the
threshold of probing new physics. There were studies of sneutrino
resonance production and decay at next generation $e^+e^-$ linear
colliders~\cite{snu3_ILC} and at hadron colliders by both
theoretical discussion~\cite{snu3_Tevatron-1, snu3_Tevatron-2,
snu3_Tevatron-3} and experimental measurements at the
Tevatron~\cite{rpvemu_CDF, rpvemu_D0}, via lepton flavor violating
$\hat{L}\hat{L}\hat{E}$ and $\hat{L}\hat{Q}\hat{D}$ interactions.
For single squark production, stringently constrained by indirect
low energy nucleon experiments, the baryon number violating
$\hat{U}\hat{D}\hat{D}$ couplings are negligibly small, for example
$\lambda^{''}_{11k}$ are less than $10^{-7}$ given by
nucleon-antinucleon oscillation measurements, and thus mechanics of
RPV squark resonance production at TeV hadron colliders are highly
suppressed. On the other hand, at the proposed Large Hadron electron
Collider (LHeC)~\cite{LHeC}, which provides complement to the LHC by
using the existing 7 TeV proton beam, single squark can be produced
and detected via $\hat{L}\hat{Q}\hat{D}$ couplings in the next
generation of electron-proton $e^{-}p$ collision experiments. In
this paper we investigate the potential of searching stop quark via
$e^{-}+p\rightarrow \tilde{t}_1^{*} \rightarrow \mu^{-} + \bar{b} $
resonance process, which provides a new prospect to probe the RPV
lepton flavor violating interactions.

\vskip 5mm
\section{Signal and Background at the $LHeC$}
\begin{figure}[htbp]
\vspace*{-0.3cm} \centering
\includegraphics[scale=0.3]{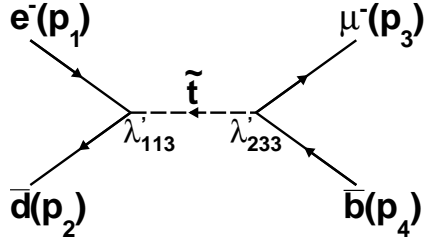}
\vspace*{-0.3cm} \centering \caption{The parton-level Feynman
diagram of RPV signal $e^{-}\bar{d}\rightarrow \mu^{-}\bar{b}$.}
\label{fig:feynman}
\end{figure}

Under the single dominance hypothesis \cite{rpv_report} that
$\tilde{t}_1$, the lighter mass eigenstate of the two stop quarks,
is simply governed by $\hat{L}\hat{Q}\hat{D}$ couplings
$\lambda_{131}^{'}$ and $\lambda_{233}^{'}$, the parton-level signal
process can be denoted as $e^-(p_1) + \bar{d}(p_2) \rightarrow
\tilde{t}_1^{*} \rightarrow \mu^{-}(p_3) + \bar{b}(p_4)$, depicted
by the Feynman diagram in FIG.~\ref{fig:feynman}.
\par
The amplitude of the signal process at parton-level can be written
as
\begin{equation}
\label{eqn:M-RPVstop} {\cal M} =
\bar{\mbox{v}}(p_2)\left[\lambda_{131}^{'}
\frac{1-\gamma^5}{2}\right]\mbox{u}(p_{1}) \cdot
\frac{-i}{\hat{s}-M^{2}+iM\Gamma } \cdot \bar{\mbox{u}}(p_3)
\left[\lambda_{233}^{'} \frac{1-\gamma^5}{2}\right] \mbox{v}(p_4)
\end{equation}
where $\sqrt{\hat{s}}=M_{\mu b}$ is the center-of-mass energy of the
hard scattering and equivalent to the final state invariant mass.
The parameter $M$ and $\Gamma$ denote the mass and total width of
the lighter stop quark $\tilde{t}_1$ respectively, while the lighter
stop is assumed only decaying through $ed$ and $\mu b$ modes.
\begin{equation}
\label{eqn:G-RPVstop} \Gamma = \frac{{\lambda_{233}^{'}}^{2}}{16\pi }
\cdot \frac{{(M^{2}-m_{b}^{2})}^{2}}{M{(M^{2}+m_{b}^{2})}} +
\frac{{\lambda_{131}^{'}}^{2}}{16\pi } \cdot M
\end{equation}
The parton-level differential cross section for signal in the rest
frame of final muon and b-quark states can be written as
\begin{equation}
\label{eqn:DeltaSigma}
\frac{d\hat{\sigma}}{d\Omega}=\frac{(\lambda_{131}^{'}\lambda_{233}^{'})^{2}}{(16\pi)^{2}\hat{s}}
\frac{(\hat{s}-m_{b}^{2})^{2}} { (\hat{s}-M^{2})^{2}+(\Gamma M)^{2}
}
\end{equation}
For the particle level signal process $e^{-}+p\rightarrow
\tilde{t}_1^{*} \rightarrow \mu^{-} +\bar{b}$ at the LHeC, the cross
section and kinematic distributions can be obtained by convoluting
the parton-level subprocess with the parton distribution function
(PDF) of the proton.

\par
The indirect two standard deviation bounds on the coupling constants
and the mass of stop are given as~\cite{rpv_report}
\begin{equation}
\label{eqn:rpvpara} \lambda^{'}_{131}\le0.03, \hskip5mm
\lambda^{'}_{233}\le0.45, \hskip 5mm M \ge 100~GeV
\end{equation}
This set of parameter limitations will be used as default for demonstration
purpose unless explicitly stated otherwise.

\par
\begin{figure}
\vspace*{-0.3cm} \centering
  \includegraphics[scale=0.7]{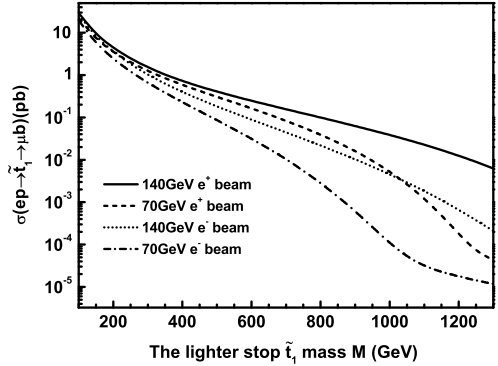}
\vspace*{-0.3cm} \centering  \caption{\label{fig:sigxsect} The cross
sections for stop resonance production $\sigma(ep\rightarrow
\tilde{t}_1
  \rightarrow \mu b)$ at the LHeC as functions of stop mass.}
\end{figure}

There exist options of lepton beam configuration, namely electron or
positron beam with 70 or 140 GeV energy. Positron beam
configurations are taken into account as comparisons with their
charge-conjugated electron beams. The signal cross sections
evaluated under different beams as a function of stop quark mass are
depicted in FIG.~\ref{fig:sigxsect}.
The positron beam options raise larger cross sections for the signal
$e^{+}+d\rightarrow\tilde{t}_1\rightarrow \mu^{+}+ b$ than those for
its charged-conjugated process in most mass region, simply because
of large density of high energy valance d-quark in proton PDF. To have our
discussion conservative, we take 70 GeV electron beam
as default except stating explicitly otherwise.

\par
To simulate the kinematics of the RPV signal and SM predictions, the
{\sc {comphep}}\cite{CompHep} event generator and {\sc
{cteq6l1}}\cite{CTEQ6-1, CTEQ6-1} PDF are used. The reducible SM
background in electron beam configuration comes from
$e^{-}+p\rightarrow e^{-}+b/\bar{b} \rightarrow \nu_e + b/\bar{b} +
W^{-}$, where the on-shell W boson decays leptonically via
$\mu^{-}\bar{\nu}_{\mu}$ channel. Since from experimental point of
view, it is difficult to determine the original charge of quarks in
reconstructed jets, both $b$ and $\bar{b}$-quark initial state
contributions at parton-level should be taken into account as
background to the muon and b-jet associated signal. However, a real
(virtual) top-quark could be produced via $e^{-}+\bar{b}\rightarrow
\nu_e + \bar{t}^{(*)}$, and enhance the cross section greatly in the
$\bar{b}$-quark channel against $b$-quark contribution, i.e.
$\hat{\sigma}(e^{-}+\bar{b}\rightarrow \nu_e+W^{-}+\bar{b})$ is
about two order greater in magnitude than
$\hat{\sigma}(e^{-}+b\rightarrow \nu_e +W^{-}+b)$ in most kinematic
region. Therefore, we only choose $e^{-}+p\rightarrow \nu_e+W^{-}
+\bar{b}$ in electron beam and its charge conjugation
$e^{+}+p\rightarrow \bar{\nu}_e+ W^{+} +b$ in positron beam
configuration as dominant SM background to $e^{\pm}+p\rightarrow
\mu^{\pm}+b/\bar{b}$ signal. In the numerical calculation we
take $m_e=0.511~MeV$,
$m_{\mu}=105.658~MeV$, $m_W=80.365~GeV$, $m_t=173.5~GeV$,
$m_b=4.65~GeV$, $\Gamma_{total}^t=2.0~GeV$,
$\alpha_{ew}^{-1}=137.036$, and set the factorization scales for
signal and background processes as $\mu_f=M$ and $\mu_f=m_t$,
respectively.

\par
\begin{figure}
\vspace*{-0.3cm} \centering
\includegraphics[scale=0.7]{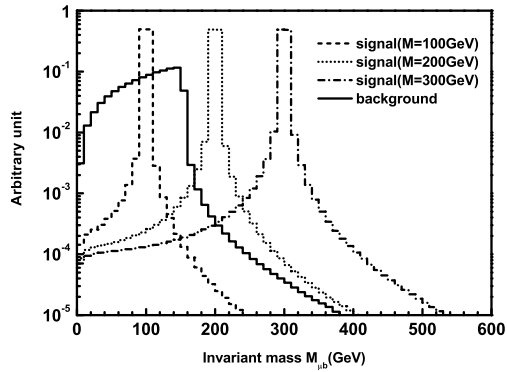}
\vspace*{-0.3cm} \centering  \caption{\label{fig:compMmb} The
invariant mass $M_{\mu b}$ distributions of background and signals
with 70GeV electron beam at the LHeC.}
\end{figure}
\begin{figure}
\vspace*{-0.3cm} \centering
\includegraphics[scale=0.7]{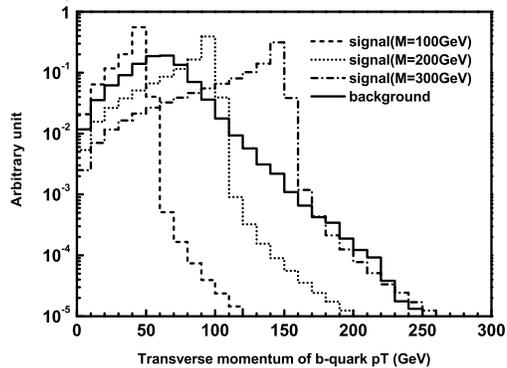}
\vspace*{-0.3cm} \centering  \caption{\label{fig:compPTb} The
b-quark transverse momentum distributions of background and signals
with 70GeV electron beam at the LHeC.}
\end{figure}
\begin{figure}[htbp]
\vspace*{-0.3cm} \centering
\includegraphics[scale=0.65]{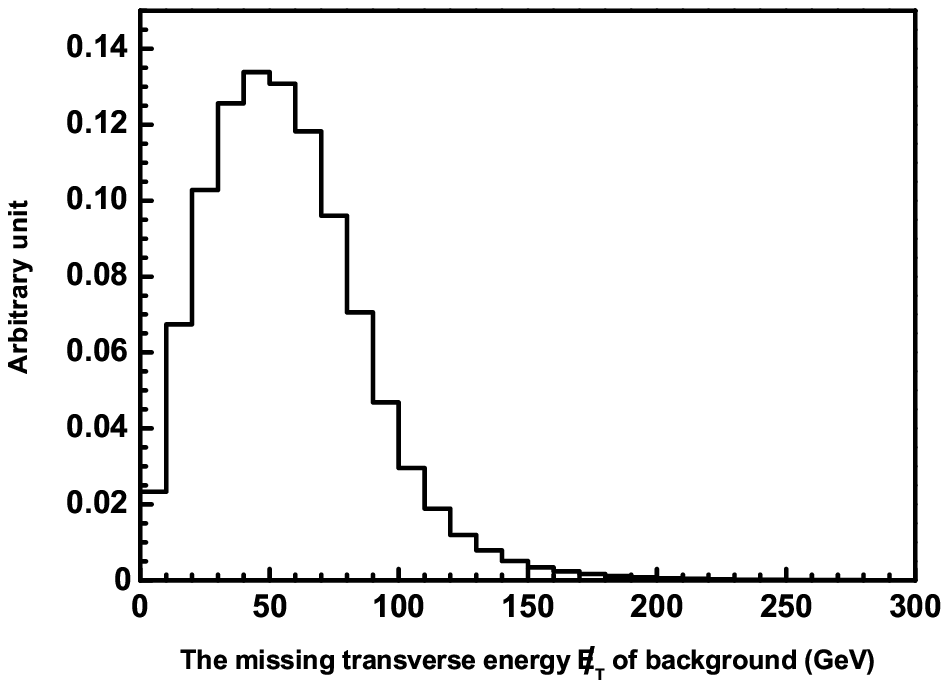}
\includegraphics[scale=0.65]{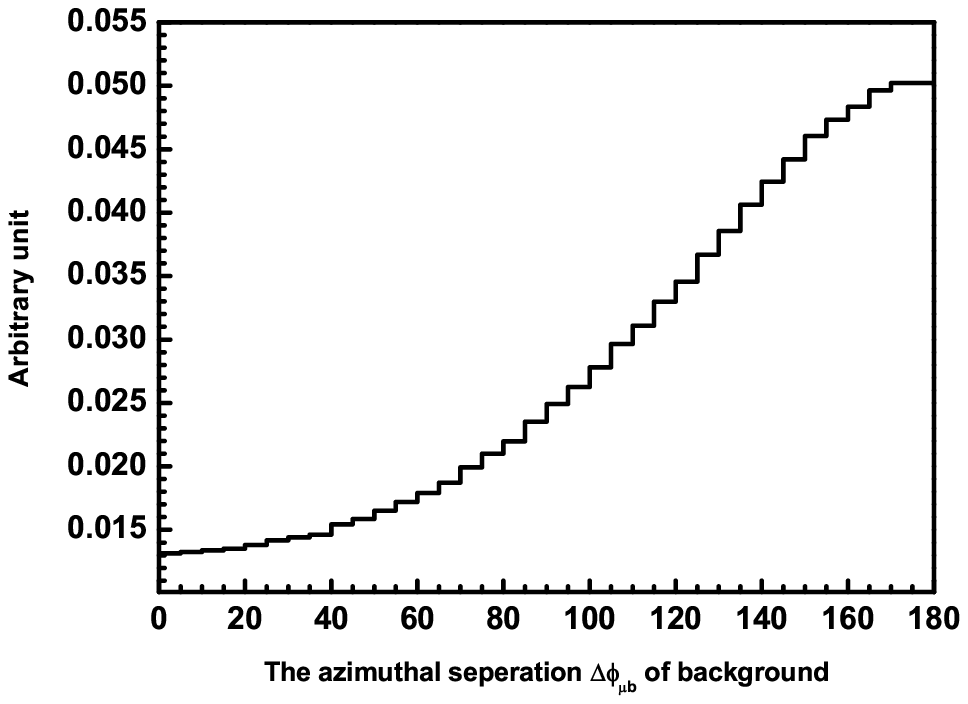}
\vspace*{-0.3cm}
\centering \caption{\label{fig:bgkMET} The distributions of
$\slashed{E}_T$ and the azimuthal separation $\Delta\phi_{\mu b}$ of
background with 70GeV electron beam at the LHeC.}
\end{figure}

In SM background, the two neutrinos $\nu_e$ and $\nu_{\mu}$ will
escape detection, and result in significant missing transverse
energy $\slashed{E}_T$. The remaining muon and b-quark are relatively
soft, and would mimic the $\mu+b$ signal. On the other hand,
governed by RPV Yukawa couplings, the outgoing muon and b-quark in
signal are isotropic in the rest frame of stop quark, and therefore
the transverse momentum $p_T$ of the two final state particles are
inclined to be as hard as taking half of the stop mass. Accordingly,
the signal is characterized by isolated and high $p_T$ outgoing muon
and b-jet, back-to-back in the transverse $r-\phi$ plane, without
sensible missing transverse energy. In experimental point of view,
these features will help to distinguish signal from background. The
comparisons of kinematic distributions of background and signals at
the LHeC are given in FIG.~\ref{fig:compMmb} and
FIG.~\ref{fig:compPTb}.
The distributions of the non-zero $\slashed{E}_T$ and azimuthal
separation of final muon and b-quark of background are shown in
FIG.~\ref{fig:bgkMET}.

\vskip 5mm
\section{Event Selection and Discussion}
Taking advantage of distinction between the RPV SUSY and SM
predictions on $\mu+b+X$ final states, a strategy of event selection
can be developed as
\begin{itemize}
\item Kinematics cuts: for muon, $p_T^{\mu}>25$GeV,
$|\eta_{\mu}|<$2.5; for b-jet, $p_T^{b}>25$GeV, $|\eta_{b}|<$3.5.

\item The open angle in transverse plane $\Delta\phi_{\mu b} >2.94$,
in which the deviation from exact back-to-back of the muon and b-jet
leaves room for the effect of initial state radiation and
resummation in signal.

\item The missing transverse energy veto as $\slashed{E}_T <25$ GeV,
taking possible energy resolution into account.

\item The invariant mass $M_{\mu b}>85$ GeV. This loose mass cut is to
have the most efficiency of 100GeV stop, assuming 3$\sigma$ b-quark
jet energy resolution below the resonance mass pole, and the cut
will not be shifted for other signal mass point to derive most
conservative estimation.

\item A 60\% b-tagging efficiency is assumed for b-jet
identification in experiments.
\end{itemize}

\par
If we assume the b-tagging efficiency is $100\%$, these criteria could suppress the SM
background efficiently. Only about $5.85\%$ $\nu_e bW(\mu\nu_{\mu})$ events will survive after
taking above selection rules, which corresponds to a cross section of $9.65~fb$.
On the other hand, $86\%$ signal events survive the above selections at stop mass $M=100~GeV$,
and the efficiency of signal would rise with the increment of stop mass from $100~GeV$ to $300~GeV$. Assuming
there is no apparent deviation from SM prediction of $\mu+b$ final
states, the 2$\sigma$ exclusion limits on signal can be derived, by
simply using significance method $\frac{S}{\sqrt{S+B}}\le 2.0$.
TABLE ~\ref{cutflow} gives the cross sections of signal after event
selection, and the minimum luminosity needed to draw 2$\sigma$
exclusion at RPV couplings given by Eq.(\ref{eqn:rpvpara}), with
$e^{\pm}p$ collider options respectively.
\begin{table}[htb]
  \centering
  \begin{tabular}{|c|c|c|c|c|c|}
  \hline
    $M$  &$\sigma(e^+p)$  & exclusion ${\cal L}$$(e^+p)$  &$\sigma(e^-p)$  & exclusion ${\cal L}$$(e^-p)$\\
    $(GeV)$ & $(pb)$         &$(pb^{-1})$           & $(pb)$         &$(pb^{-1})$\\

  \hline
    100 & $19.15$  & $0.348$     & $16.80$           & $0.40$\\
  \hline
    200 & $3.45$   & $1.94$     & $2.37$         & $2.81$\\
  \hline
    300 & $1.21$   & $5.54$     & $0.63$         & $10.69$\\
  \hline
    400 & $0.56$   & $12.19$      & $0.22$       & $32.31$\\
  \hline
    500 & $0.28$   & $24.64$      & $7.82\times 10^{-2}$  & $95.79$\\
  \hline
    600 & $0.14$   & $50.03$      & $2.73\times 10^{-2}$  & $330.43$\\
  \hline
    700 & $6.94\times 10^{-2}$    & $109.36$    & $8.52\times 10^{-3}$  & $1.69\times 10^{3}$\\
  \hline
   800  & $3.10\times 10^{-2}$    & $282.27$    & $2.22\times 10^{-3}$  & $1.61\times 10^{4}$\\
  \hline \end{tabular}
  \caption{\label{cutflow} \small
    The cross sections and the minimal luminosities of 2$\sigma$ exclusion at the LHeC with $70~GeV$ $e^{\pm}$ beam.
  }
\end{table}

\par
\begin{figure}[htbp]
\vspace*{-0.3cm} \centering
\includegraphics[scale=0.7]{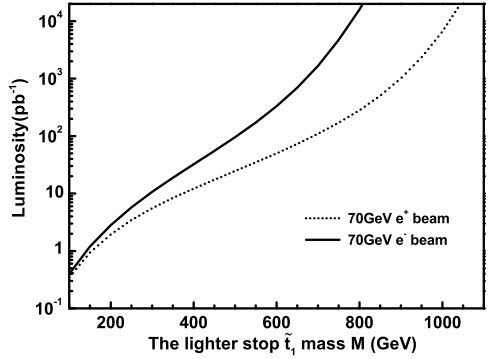}
\vspace*{-0.3cm}
\centering  \caption{\label{fig:LumiMass} The expected luminosity
for 2$\sigma$ exclusion of RPV signals at the LHeC.}
\end{figure}

The luminosities required to exclude stop $\mu+b$ signal at a 70
$GeV$ electron- or positron-proton LHeC collider are also depicted
in FIG.~\ref{fig:LumiMass}.
The obvious better sensitivity of positron beam in probing high mass
RPV signal than that of electron beam, is simple raised by the
larger parton density of valance $d$-quark than that of sea
$\bar{d}$-quark in proton. One can see that with 1$fb^{-1}$ at $70~GeV$ $e^-p$ collider, the
RPV $\mu+b$ resonance of stop quark can be excluded up to about
$M\le 700~GeV$ with default RPV couplings.

\par
\begin{figure}[htbp]
\vspace*{-0.3cm} \centering
\includegraphics[scale=0.7]{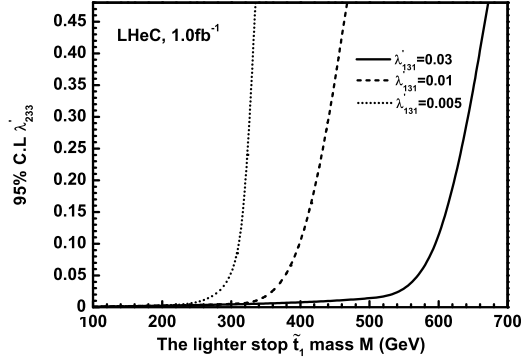}
\vspace*{-0.3cm}
\centering  \caption{\label{fig:CoupMass}The upper bounds on
$\lambda_{233}^{'}$ at given $\lambda_{131}^{'}$ as functions of
stop masses at 70GeV $e^-p$ collider.}
\end{figure}

New direct 2$\sigma$ upper bounds on $\lambda_{233}^{'}$ at given
$\lambda_{131}^{'}$ can be calculated as a function of stop mass,
and are depicted at 70GeV electron beam case in
FIG.\ref{fig:CoupMass}.
More stringent constraints on RPV couplings could be given at LHeC.
For example, at energy scale about $100~GeV$, constraints as
$\lambda^{'}_{131}=0.005$, $\lambda^{'}_{233} \lesssim 0.85\times 10^{-3}$ can be derived
from exclusive $\mu+b$ resonance search, which improve the
associated sensitivity on $\lambda^{'}_{131}\times
\lambda^{'}_{233}$ by nearly 4 orders compared to those given in
Eq.(\ref{eqn:rpvpara}) from low energy processes. Moreover, in high
mass region up to 500GeV, it is likely that either the stop RPV
$\mu+b$ resonance could be directly detected, or
$\hat{L}\hat{Q}\hat{D}$ couplings would be constrained much more
stringently at the LHeC than given by indirect low energy
experiments.

\par
Some issues should be addressed here:
\begin{itemize}
  \item
  The purpose of this study is to demonstrate the potential of the LHeC
  experiments in probing RPV single squark resonance with even the
  most conservative estimations. Thus, the event selection strategy
  developed above is practicable and moderate, which is far from strict
  enough to optimize the signal significance. The constraints
  on R-violating $\hat{L}\hat{Q}\hat{D}$ couplings derived are conservative and easy to
  achieve in experiment. For example, more severe moun b-jet mass window
  cut or binned likelihood method on $M_{\mu b}$ can be employed while
  searching signal in high mass region, which would significantly improve
  sensitivity for $M>500$GeV easily.

  \item
  The large signal cross section of positron configuration $e^{+}p$ over electron beam $e^{-}p$ in
  direct searching stop $\tilde{t}_1$ quark resonance, is simply due to
  the large density of valance d-quark in proton. On the other hand,
  electron beams will take advantage over much larger luminosity; moreover, the single sbottom
  quark $\tilde{b}$ resonance production and decay at the LHeC,
  i.e. $e^{-} + p \rightarrow \tilde{b}\rightarrow \mu^{-} + u_{k}$
  analogically could be dominant. Therefore, the electron beam configuration can
  provide excellent opportunity to probe $\lambda^{'}_{113}$ and $\lambda^{'}_{2k3}$
  interactions.
\end{itemize}

\vskip 5mm
\section{Summary}
In this paper, the possibility of probing lepton flavor changing RPV
$\hat{L}\hat{Q}\hat{D}$ interactions via $e+p\rightarrow \tilde{t}
\rightarrow \mu+b$ process at the LHeC collider is investigated.
Under the single dominance hypothesis, the resonance of stop quark
can be produced and dominantly decay into muon and b-quark final
states. An event selection strategy is developed to optimize the
sensitivity of signal over SM background. Taking advantage of the
enhancement of the direct resonance production of squark and the
distinctive kinematics distributions between the signal and SM
predictions, we come to conclusions that if there is no apparent
excess of SM predictions on $\mu+b$ final states, the sensitivity of
RPV interactions can be measured at an unprecedented level compared
to all the knowledge derived from indirect measurements.

\vskip 5mm
\par
\noindent{\large\bf Acknowledgments:} This work was supported in
part by the National Natural Science Foundation of China
(No.11075150, No.11025528, No.11005101) and the Specialized Research
Fund for the Doctoral Program of Higher Education
(No.20093402110030).

\end{document}